\documentclass[useAMS,usenatbib,rotating]{mnras}
\usepackage{amssymb}     \usepackage{amsmath}    \usepackage{amsfonts}
\usepackage{xspace}    \usepackage{longtable}    \usepackage{rotating}
\usepackage{lscape}
\usepackage{times}

\newcommand{\HI}{\mbox{\sc      H{i}}}     
\newcommand{\Htwo}{\mbox{\sc H$_2$}}
            
\newcommand{\HII}{\mbox{\sc            H{ii}}}

\newcommand{\CO}{\mbox{$^{12}$CO(1-0)}}
\newcommand{\msun}{\mbox{$M_\odot$}}    
              
\newcommand{\kms}{\mbox{km    s$^{-1}$}}

\newcommand{\lsun}{\mbox{$L_\odot$}}

\title[LMT discovery of $^{12}$CO(1-0) in VIIZw466]
  {Early Science with the Large Millimeter Telescope: discovery of the $^{12}$CO(1-0) emission line in the ring galaxy, VIIZw466}
\author[O.\ I.\ Wong et al.]
  {O. Ivy Wong,$^{1,2}$\thanks{E-mail:ivy.wong@uwa.edu.au} O.~Vega,$^{3}$  D. S\'anchez-Arg\"uelles,$^{3}$  G.~Narayanan,$^{4}$ W.F.~Wall,$^{3}$
\newauthor M.A.~Zwaan,$^{5}$ D.~Rosa~Gonz\'alez,$^{3}$ M.~Zeballos,$^{3}$ K.~Bekki,$^{1}$ Y.D.~Mayya,$^{3}$ \newauthor A.~Monta\~na,$^{3,6}$  \& A. Chung$^7$\\
  $^1$ International Centre for Radio Astronomy Research, University of Western Australia M468, 35 Stirling Highway, Crawley, WA 6009, Australia\\
  $^2$ ARC Centre of Excellence for All-Sky Astrophysics (CAASTRO), Australia\\
  $^3$ Instituto Nacional de Astrof\'isica, Optica y Electr\'onica, Luis Enrique Erro 1, Tonantzintla, Puebla, M\'exico\\
  $^4$ Department of Astronomy, University of Massachusetts, Amherst, MA 01003, USA \\ 
  $^5$ European Southern Observatory, Karl-Schwarzschild-Str. 2, D-85748 Garching bei M\"unchen, Germany \\
  $^6$Consejo Nacional de Ciencia y Tecnología, Av. Insurgentes Sur 1582, Col. Cr\'edito Constructor, Del. Benito Ju\'arez, C.P.: 03940, D.F., M\'exico \\
  $^7$ Department of Astronomy \& Yonsei University Observatory, Yonsei University, 50 Yonsei-ro, Seodaemun-gu, Seoul 120-749, Korea\\
 }
\date{Released 2016 Xxxxx XX}

\pagerange{\pageref{firstpage}--\pageref{lastpage}} \pubyear{2016}

\def\LaTeX{L\kern-.36em\raise.3ex\hbox{a}\kern-.15em
    T\kern-.1667em\lower.7ex\hbox{E}\kern-.125emX}

\begin{document}

\label{firstpage}

\maketitle

\begin{abstract}
We report an early science discovery of the \CO\  emission line in the
collisional ring galaxy, VII~Zw466, using the Redshift Search Receiver instrument
 on the Large Millimeter Telescope Alfonso Serrano.
The apparent molecular-to-atomic gas ratio either places the ISM of VII~Zw466 in 
the \HI-dominated regime or implies a large quantity of CO-dark molecular gas, 
given its high star formation rate. The molecular gas densities and 
star formation rate densities of VII~Zw466 are consistent with the standard 
Kennicutt-Schmidt star formation law even though we find this galaxy to be \Htwo-deficient.
The choice of CO-to-\Htwo\ conversion factors cannot explain the apparent \Htwo\
 deficiency in its entirety. Hence, we find that the collisional ring galaxy, VII~Zw466, 
is either largely deficient in both \Htwo\ and \HI\ or contains a large mass of 
CO-dark gas. A low molecular gas fraction could be due to the enhancement of feedback 
processes from previous episodes of star formation as a result of the star-forming 
ISM being confined to the ring.   We conclude that collisional ring galaxy 
formation is an extreme form of galaxy interaction that triggers a strong galactic-wide 
burst of star formation that may provide immediate negative feedback towards subsequent
 episodes of star formation---resulting in a short-lived star formation history or, at least,
 the appearance of a molecular gas deficit.

\end{abstract}


\section{Introduction}
Drop-through or collisional ring galaxies are unique astrophysical objects in that 
the physical mechanism that drives the ring formation --- that of expanding 
density waves---  are very well-understood. 
The process of star formation plays a vital role in the evolution of galaxies.
In collisional ring galaxies, the star formation within the ring is a result of 
the interaction-triggered density waves which compress the cold
gas in the galaxy's interstellar medium (ISM) \citep[e.g.\ ][]{lynds76,wong06}.

Similar to the prototypical collisional ring galaxy, the Cartwheel galaxy 
\citep[A0035-33; e.g.\ ][]{appleton96, higdon15}, VII~Zw466 is a collisional ring galaxy in a 
compact galaxy group with two other massive companion galaxies (see 
Figure~\ref{pointing}).  However, this is where the similarity ends.  VII~Zw466 
has been classified as an empty ring (RE) galaxy and the result of an off-centre 
collision \citep{mapelli12,fiacconi12}, while the Cartwheel is likely a product of a 
nuclear collision \citep{appleton97}.  Also, observations have found VII~Zw466
 to consist of an ISM with solar metallicities \citep{bransford98}.  Hence,
the combination of the ring formation process and the metallicity suggests 
that VII~Zw466 is more similar to AM0644-741 \citep{higdon11} and NGC~922
\citep{wong06}.  An \HI\ bridge connects VII~Zw466 and its southern companion, G2 
\citep[see Figure~1; ][]{appleton96}.  Table~\ref{genprops} presents the general 
optical and \HI\ properties  of VII~Zw466 \citep{romano08, appleton96}.

Galaxy interactions can increase the density of \Htwo\ molecular gas and lead 
to strong bursts of star formation \citep[e.g.\ ][]{henderson16}.  However,
previous CO observations have found collisional ring galaxies to be relatively 
deficient in molecular gas \citep[e.g.\ ][]{horellou95}.  More recently, \citet{higdon11} 
has found evidence for enhancement in the rate of photodissociation of 
the molecular \Htwo\ in these galaxies.  That is, the destruction of \CO\ appears to be more 
efficient relative to the molecular clouds within the star-forming disks of other
 field galaxies whose star formation is not triggered by the compression of the
 ISM \citep{higdon11}.  Such extreme `over-cooked' ISM is likely to be 
dominated by small clouds and results in the underestimation of \Htwo\ 
 from \CO\ observations when using the standard $X_{\rm{CO}}$ conversion \citep{higdon11}.

Typically, $N(H_2) = X_{\rm{CO}} S_{\rm{CO}}$ where $N(H_2)$ is the \Htwo\ column density in
 units of cm$^{-2}$ and $S_{\rm{CO}}$ is the integrated \CO\ line intensity in K~km~s$^{-1}$
 \citep[e.g.\ ][]{bolatto13}.  The alternative \CO-to-\Htwo\ conversion factor is 
$\alpha_{\rm{CO}}$ where
the molecular gas mass, $M_{\rm{MOL}} = \alpha_{\rm{CO}} \times L_{\rm{CO}}$  \citep{bolatto13} 
where $L_{\rm{CO}} = 2453\, S_{\rm{CO}} \Delta v D_{\rm{L}}^2 / (1+z)$ \citep{solomon05} in
units of K~km~s$^{-1}$~pc$^2$.
The $X_{\rm{CO}}$ can be obtained by multiplying $\alpha_{\rm{CO}}$ by $4.6 \times 10^{19}$
 \citep[e.g.\ ][]{sandstrom13}.  The $\alpha_{\rm{CO}}$ conversion factor 
includes a factor of 1.36 for helium.

Current theory suggests that high surface density regions (with greater gas 
temperatures and velocity dispersions) have a decreased $X_{\rm{CO}}$ value relative 
to that of our Galaxy \citep{narayanan11,narayanan12}.  However, the accuracy of 
\CO\ as a tracer of the total \Htwo\ in low metallicity galaxies is more complex
 and difficult to determine.  An increase in $X_{\rm{CO}}$
 can be largely attributed to a contraction of the emitting surface area relative
to that of the \Htwo\ \citep{bolatto13}.  In addition, this increase in $X_{\rm{CO}}$ may be further
 compounded or offset by a variation in brightness temperature due to variations
in the heating sources (see Section 6.1 of \citet{bolatto13} for a more detailed discussion).
Therefore, despite recent advances in theoretical modelling \citep[e.g.\ ][]{shetty11,narayanan11,narayanan12}, 
the $X_{\rm{CO}}$  factor remains an area of significant uncertainty in the 
study of \Htwo\ in collisional ring galaxies.

\begin{table}
\caption{General properties of VII~Zw466 from \citet{romano08}, \citet{appleton96} and \citet{devaucouleurs91}.}
\label{genprops}
\footnotesize{
\begin{center}
\begin{tabular}{lc}
\hline
\hline
Property  &  \\
\hline
Right ascension (J2000) &  12:32:04.4 \\ 
Declination (J2000) & +66:24:16 \\
Distance ($D_L$) & 207 Mpc \\
Heliocentric velocity & 14,490 \kms\  \\
Ring diameter & 21.8 kpc\\
Stellar mass &  $4.83 \times 10^{10}$ \msun\ \\
\HI\ mass & $4.1 \times 10^9$ \msun\ \\
SFR & 7.8 \msun\ year$^{-1}$ \\
\hline
\hline
\end{tabular}
\end{center}}
\end{table}

To further investigate the molecular gas properties in the potentially `overcooked' 
ISM conditions of collisional ring galaxies, we search for \CO\ in the ring 
galaxy VII~Zw466.  
This galaxy is an excellent candidate for such an investigation because of the following: 
1) we can discount the effects of metallicity, and 2) we have already assembled an 
excellent repository of multiwavelength observations such as the \HI, H$\alpha$
and the far-infrared (FIR) to constrain the atomic gas and star formation 
properties of this galaxy.  

A previous attempt to measure the molecular gas content of VII Zw466 has resulted 
only in upper limits \citep{appleton99}.  The molecular component
is the least-explored  but one of the most crucial properties required to 
understand star formation.  Therefore, the aims of this
paper are as follows: 1) to determine the molecular gas content of VII Zw466 using the
Large Millimeter Telescope Alfonso Serrano \citep[LMT; ][]{hughes10}, and 2) to
 determine the extent to which a variation of the $X_{\rm{CO}}$ value can
 affect our estimate of the total \Htwo\ mass.

The details of our observations and processing can be found in Section 2. 
Section 3 presents the results of our observations.  We discuss the implications
of our results in Section 4 and Section 5 presents a summary of our findings.
Throughout this paper we adopt a $\Lambda$CDM cosmology of $\Omega_m = 0.27$, 
$\Omega_{\Lambda}=0.73$ with a Hubble constant of $H_{\rm{0}}=73$~km~s$^{-1}$~Mpc$^{-1}$.

\section{Observations and processing}
\subsection{Observations}
The \CO\ observations of VII~Zw466 are obtained using the 
 Large Millimeter Telescope ``Alfonso Serrano'' \citep[LMT; ][]{hughes10}  and the
Redshift Search Receiver \citep[RSR; ][]{erickson07} that are being commissioned 
at the peak of Sierra Negra, Mexico.  Our observations are obtained in June 2015
as part of the Early Science programme when only the inner 32~m of the 50~m 
primary reflector is fully operational.  Upon completion, the LMT 50~m will 
become the  world's largest dedicated millimeter single dish telescope, providing 
good spatial resolution and sensitivity to extended emission.

The  RSR \citep{erickson07,chung09,snell11}   is the ideal instrument for the 
purposes of this project as it possesses an 
ultra-wide bandwidth that spans the frequencies of 73--111~GHz with  spectral 
channel widths of 31~MHz (corresponding to $100$~\kms at 93~GHz).
 Such a wide bandwidth maximises the number of molecular  lines that can 
be observed in a single snapshot.  The RSR beamsize ranges from 20~arcseconds to 28~arcseconds.
We expect a beamsize of $\sim20$~arcseconds for the \CO\ observations presented 
in this paper.  

\begin{figure}
\begin{center}
\includegraphics[scale=.5]{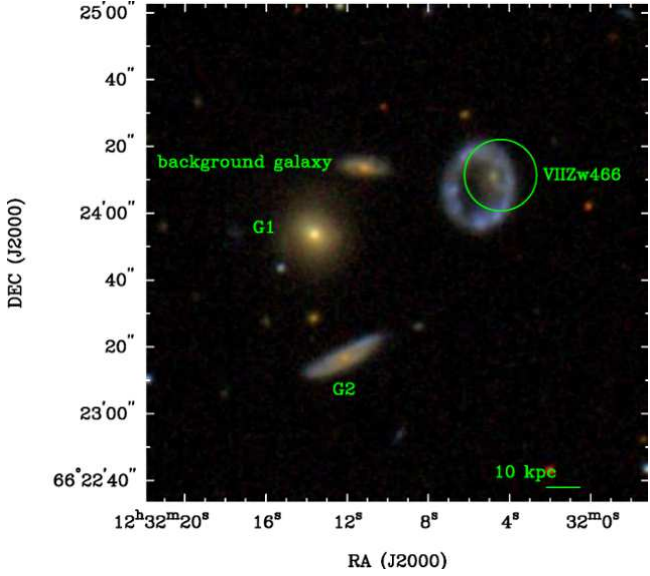} 
\end{center}
\caption{Three-colour SDSS DR12 \citep{alam15} map of VII~Zw466 and its two neighbouring galaxies, labelled `G1' and `G2'. Blue: $g$-band; green: $r$-band;and red: $i$-band.  The green circle marks a representative 20-arcsecond beam centred on the LMT pointing described in this paper.   The southern companion galaxy G2 is thought to be the intruder galaxy that triggered the formation of the star-forming ring in VII~Zw466 \citep{appleton96,romano08}. An \HI\ bridge connects VII~Zw466 and G2 \citep{appleton96}. The scale bar at the bottom right corner represents 10~kpc.}
\label{pointing}
\end{figure}

We observed VII~Zw466 in a single pointing for a total integration time of 
275~minutes over 4 nights (7--19 June~2015) in average weather conditions 
where $T_{\rm{sys}}\sim 119$~K.  Figure~\ref{pointing} shows the location of this 
single pointing overlaid on the three-colour map of VII~Zw466 from the Sloan
Digital Sky Survey \citep[SDSS; ][]{alam15}.

The RSR has two beams that are used to observe the source and the sky simultaneously
 in order to account for the atmospheric contribution to our observations. 
Our observations were conducted using the ``wide beam throw'' observing mode (where 
the two beams are separated by 294~arcseconds).  The telescope pointing calibrator, 
QSO~1153+495, was used.

\begin{figure}
\begin{center}
\includegraphics[scale=.47]{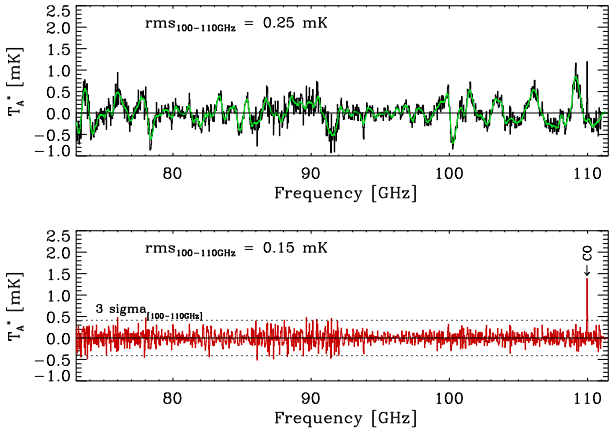} 
\end{center}
\caption{Fitting the calibrated RSR spectrum with a Savitzky-Golay 
(S-G) filter in order to account for the artefacts present in our bandpass.  
Top panel: the calibrated RSR spectrum overlaid with the fitted S-G filter 
(shown in green).  Bottom panel: the final spectrum after the S-G filter has
been subtracted. The dotted lines represent the $\pm3$-$\sigma$ level of this spectrum.  
The small arrows mark the location of the redshifted \CO\ 
emission line.  The brightness scale of the above spectra is that of $T_{\rm{A}}^*$.}
\label{reducedrsr}
\end{figure}

\subsection{Processing}
Our observations are calibrated and processed using the latest version of  
{\sc{Dreampy}}, a software package written by G.\ Narayanan for the purposes of 
calibrating and reducing LMT--RSR observations. As the entire RSR 38~GHz bandwidth
is distributed across six spectrometer boards, each individual spectrum is 
processed independently. A linear baseline is fitted and removed from each spectrum
 after the removal of bad channels and/or spectra containing strong baseline ripples.
The resulting 38~GHz spectrum is produced by co-adding all the individual spectra
 with a 1/$\sigma^2$ weighting.   We show the reduced spectrum that has been
 corrected for atmospheric losses, rear spillover and scattering in Figure~2  on 
the $T_{\rm{A}}^*$ scale.

Even though the RSR system and the {\sc{Dreampy}} reduction 
pipeline are optimised to improve the stability in the spectral baseline over 
the entire frequency range, some wide bandwidth artefacts remain (top panel of
Figure~\ref{reducedrsr}).  To  reduce the effects of the aperiodic ripples that
 remain in the reduced spectrum, we applied a Savitzky-Golay (S-G) filter to
 the entire spectrum \citep{cybulski16}.
We used an S-G filter with a 2nd order polynomial fit with a width of 1~GHz.
It should be noted that 1~GHz is significantly greater than the width of 
any expected molecular line in our spectrum.  The bottom panel of 
Figure~\ref{reducedrsr} shows the resulting spectrum after the S-G filter
has been applied.  The RMS of the spectrum between 100~GHz and 111~GHz has 
decreased from 0.25~mK to 0.15~mK after the application of the S-G filter.

\section{Results}

\subsection{Integrated \CO\ intensity}
\begin{figure}
\begin{center}
\includegraphics[scale=.47]{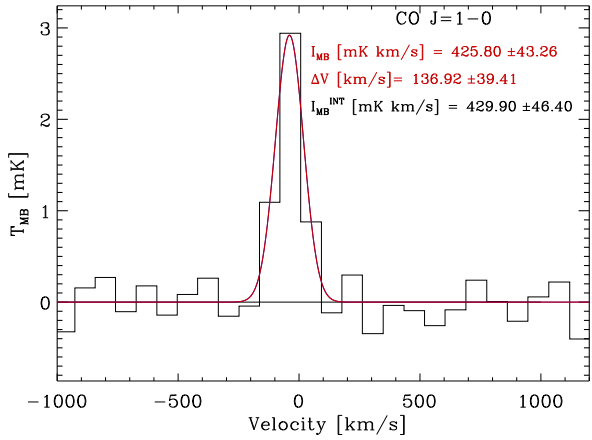} 
\end{center}
\caption{A Gaussian fit (red line) to the \CO\ emission line from VII~Zw466 on the 
$T_{\rm{MB}}$ scale. We assume the systemic velocity of VII~Zw466 to be the heliocentric velocity
(14,490 \kms) listed in Table~1.}
\label{linefit}
\end{figure}

The redshifted \CO\  emission line at 109.96~GHz is the only 
 emission line found in our RSR observations of VII~Zw466 (bottom panel of Figure~\ref{reducedrsr}).  
To convert to the main beam temperature scale ($T_{\rm{MB}}$), we estimate the 
efficiency factor to be  0.47 in the frequency range of 100--111~GHz.
Figure~\ref{linefit} presents our detection of this \CO\ emission line (on the $T_{\rm{MB}}$ scale)
 with a signal-to-noise of $\sim 10$, centred on the assumed heliocentric
velocity listed in Table~\ref{genprops}.  
We fit the emission line profile with a Gaussian (represented by the red solid line 
in Figure~\ref{linefit}).  We find the fitted velocity 
width to be 137~($\pm 39$)~\kms\ and the fitted integrated flux to be  $0.426 \pm 0.043$~K~\kms.
As the 31~MHz channel width corresponds to 84.5~\kms\ at 109.96~GHz, we
find the  \CO\ line width to be 108~($\pm 50$)~\kms. 
Uncertainties for the fitted parameters are obtained via bootstrap resampling.  

For comparison, the integrated flux  ($S_{\rm{CO}}$) measured from the reduced spectrum is $0.430 \pm 0.046$~K~\kms.  Following \citet{solomon05}, we estimate the CO luminosity ($L_{\rm{CO}}$) of VII~Zw466 to total $1.6 \, (\pm 0.5) \times 10^8$~K~km~s$^{-1}$~pc$^{2}$.
Assuming a uniform distribution of \CO\ across the entire galaxy, we estimate a
total $S_{\rm{CO}}$ of $1.229 \pm 0.078$~K~\kms\ for VII~Zw466 since our single 20-arcsecond 
pointing covers only 35~percent of VII~Zw466.  Therefore, we deduce that the
total $L_{\rm{CO}}$ of the entire ring galaxy to be $4.6 \, (\pm 0.8) \times 10^8$~K~km~s$^{-1}$~pc$^{2}$.


Using the correlations  for weakly-and strongly-interacting galaxies from
 \citet{solomon88}, we estimate $L_{\rm{CO}}$ from  the FIR luminosity ($L_{\rm{FIR}}$; 
from the IRAS Faint Source Catalog \citep{moshir90})  to range from $1.55 \times 10^{10}$~K~km~s$^{-1}$~pc$^{2}$ 
to $1.49 \times 10^{9}$~K~km~s$^{-1}$~pc$^{2}$ for  weakly-and strongly-interacting galaxies,
respectively.  Assuming that collisional ring galaxies such as VII~Zw466 can be
 classified as a strongly-interacting star-forming galaxy, the estimated total
$L_{\rm{CO}}$ is approximately 30.9~percent of the expected $L_{\rm{CO}}$.  Therefore
our measured $L_{\rm{FIR}}$/$L_{\rm{CO}}$ fraction is  greater than that
measured in giant molecular clouds (GMC) in the Galaxy  \citep{solomon87,mooney88}. 
On the other hand, the $L_{\rm{FIR}}$/$L_{\rm{CO}}$ fraction found for VII~Zw466 is 
comparable to 11 of the 14 merging galaxies examined by \citet{solomon88}.

\subsection{\Htwo\ molecular gas in VII Zw466}

The \Htwo\ content of a galaxy provides one of the fundamental building blocks 
from which stars form.  Since cold \Htwo\ is not directly observable,
astronomers typically infer the \Htwo\ content of galaxies from observations of \CO.
Many studies have found the $X_{\rm{CO}}$ conversion factor to vary
depending on galactic environment, interaction history and metallicities 
\citep[e.g.\ ][]{scoville91,wilson95,solomon97,hinz06,bolatto08,meier10,leroy11,genzel12}.

Recently, detailed theoretical simulations by \citet{narayanan12} have modelled 
the $X_{\rm{CO}}$ conversion factors as a function of the metallicity and CO line 
intensity.  As the metallicity of VII~Zw466 is approximately solar 
\citep[$Z=1$; ][]{bransford98}, the metallicity term in Equation~10 of \citet{narayanan12} 
equals 1.0 and  Equation~10 can be rewritten as: 
\begin{equation}
X_{\rm{CO}}\; = \; \rm{min}\;[4, \, 6.75 \times \langle S_{\rm{CO}} \rangle^{-0.32}] \times 10^{20} \;\; cm^{-2}~(K \, km~s^{-1})^{-1} 
\end{equation}
where $S_{\rm{CO}}$ is the integrated \CO\ brightness on the $T_{\rm{MB}}$ scale in units of K~\kms.  The 
minimum value is adopted for  $X_{\rm{CO}}$ from the above equation because
at low values of $S_{\rm{CO}}$, $X_{\rm{CO}}$ appears to be independent of 
the galactic environment and is limited mainly by the fixed galaxy properties 
(see \citet{narayanan12} for more details).  Using Equation~1 above, we 
find $X_{\rm{CO}}$ to be approximately $3.74 \times 10^{20}$~cm$^{-2}$~(K~km~s$^{-1})^{-1}$.
 
Using $M_{\rm{H2}}= 5.5 X_{CO} (2 \times 10^{20})^{-1}  L_{CO}$~K~km~s$^{-1}$~pc$^2$ \citep{leroy09} 
and assuming $X_{\rm{CO}}=3.74 \times 10^{20}$, we find the \Htwo\ mass of VII~Zw466 to be 
$4.7 \times 10^9$~\msun.   We note that this model-derived $X_{\rm{CO}}$ is approximately
1.9 times the Solar neighbourhood value of $X_{\rm{CO}}$=$2 \times 10^{20}$~cm$^{-2}$~(K~km~s$^{-1})^{-1}$  \citep{solomon87,strong96,dame01}. Consequently, the estimated \Htwo\ mass is also greater by a factor of 1.9.

To ascertain the reliability of our derived \Htwo\ mass, we test a second prescription
for the \CO-to-\Htwo\ conversion factor. \citet{bolatto13} recommends a simple
determination for the conversion based upon the metallicity ($Z^\prime$), the average surface density
of molecular clouds ($\Sigma_{\rm{GMC}}$ in units of 100~\msun\ pc$^{-2}$) and the total stellar plus gas 
density ($\Sigma_{\rm{TOT}}$):
\begin{equation}
\alpha_{\rm{CO}} \approx 2.9 \; exp \left( \frac{+0.4}{Z^{\prime}\Sigma_{\rm{GMC}}} \right) \left( \frac{\Sigma_{\rm{TOT}}}{100 {\rm{M}}_{\odot} {\rm{pc}}^{-2}} \right)^{-\gamma}
\end{equation}
$\alpha_{\rm{CO}}$ is in units of \msun\ ~(K~km~s$^{-1}$~pc$^2$)$^{-1}$, and $\gamma \approx 0.5$ for 
 $\Sigma_{\rm{TOT}} > 100$~\msun\ ~pc$^{-2}$ \citep{bolatto13}.
  
Approximating the lower limit of the total stellar plus gas
density to be the total stellar plus \HI\ gas density, we assume that all the gas and stars
are located within the ring of VII~Zw466 --- a torus with inner and outer 
diameters of 11~kpc and  21.8~kpc, respectively \citep{romano08, appleton87}. We 
estimate a total surface area of 278.2~kpc$^2$.  This 
is clearly a rough approximation, however it suffices for the purpose of 
approximating the minimum total density of the ring.  As such, we find 
$\Sigma_{\rm{TOT}} \gtrsim  188$~\msun~pc$^{-2}$ for the ring of VII~Zw466.

Estimating $\Sigma_{\rm{GMC}}$ is difficult. While we expect high 
GMC densities due to the confinement of the majority of VII~Zw466's ISM to be within
the star-forming ring, the consequence of such ISM confinement actually leads to
the enhanced destruction of molecular clouds due to: 1) expanding \HII\ regions
 \citep{bally80,maddalena87}; and 2) cloud-cloud collisions \citep{appleton87}.
This results in smaller molecular cloud fragments and photodissociated \HI\ 
\citep[e.g.\ ][]{higdon11}.  If we assume a fiducial minimum and maximum $\Sigma_{\rm{GMC}}$
 of 40~\msun\ pc$^{-2}$ and 188~\msun\ pc$^{-2}$, we find that \citet{bolatto13}'s
prescription results in $\alpha_{\rm{CO}}$ of 5.7 and 3.3~\msun\ ~(K~km~s$^{-1}$~pc$^2$)$^{-1}$,
 respectively.  These $\alpha_{\rm{CO}}$ values correspond to  $X_{\rm{CO}}$ of 
$2.6~\times~10^{20}$~cm$^{-2}$~(K~km~s$^{-1})^{-1}$ and $1.5~\times~10^{20}$~cm$^{-2}$~(K~km~s$^{-1})^{-1}$.
Therefore, the estimated \Htwo\ mass for VII~Zw466 ranges from $1.9~\times~10^9$~\msun\ 
to $3.3 \times 10^9$~\msun\ using the  prescription from \citet{bolatto13}
in combination with the assumption that $\Sigma_{\rm{GMC}}$ ranges between  40~\msun~pc$^{-2}$ 
and 188~\msun\ pc$^{-2}$.

For a standard Milky Way $\alpha_{\rm{CO}}=4.4$~\msun\ ~(K~km~s$^{-1}$~pc$^2$)$^{-1}$ 
\citep[e.g.\ ][]{strong96,dame01}, we estimate $M_{\rm{H2}}$ to be $2.6 \times 10^9$~\msun.  
Do the extreme ISM conditions within the ring of this collisional galaxy lead to 
a variation of the  $X_{\rm{CO}}$ value as a result of reduced shielding of molecular clouds 
against feedback  effects \citep{higdon11,higdon15} ? We find from our 
estimations above that such variations in $X_{\rm{CO}}$ values are likely to be relatively 
small and may only be 
factors of 0.8 to 1.9 different to the standard Galactic value.  As such, we 
estimate that the $M_{\rm{H2}}$ of VII~Zw466 to be between $1.9 \times 10^9$~\msun\ and 
$4.7 \times 10^9$~\msun.


\section{Discussion}
\subsection{Comparing our estimated $M_{\rm{H2}}$ to previous studies and known scaling relations}

In this section, we demonstrate the extreme ISM and star formation properties  of collisional 
ring galaxies by comparing multiwavelength observations of VII~Zw466 to
 standard scaling relationships and correlations found in samples of nearby star-forming
galaxies.

A previous search for \Htwo\ in VII~Zw466 yielded an upper
limit of $<1.2 \times 10^9$~\msun\ \citep[using $X_{\rm{CO}}=2.3\times 10^{20}$~cm$^{-2}$~(K~km~s$^{-1}$)$^{-1}$; ][]{appleton99}. When we adopt the same $X_{\rm{CO}}$ value as \citet{appleton99}, we find
 a total $M_{\rm{H2}}$ of $2.9 \times 10^9$~\msun. Such a discrepancy in total $M_{\rm{H2}}$ between the two sets of observations argues strongly against our previous assumption that the \Htwo\ content is uniformly distributed across this galaxy.  As such, it is likely that the observed  $M_{\rm{H2}}$ in our single LMT pointing accounts for at least 84~percent of the total $M_{\rm{H2}}$ in VII~Zw466.  

Using the 60~$\mu$m and 100~$\mu$m observations from the IRAS Faint Source Catalog \citep{moshir90}, we estimate that the  $L_{\rm{FIR}}$ of VII~Zw466 is $2.73 \times 10^{10}$~\lsun.
From  $L_{\rm{FIR}}$, we would expect $L_{\rm{CO}}$ to be 
$8.4 \times 10^{8}$~K~km~s$^{-1}$~pc$^{2}$ and $1.6 \times 10^8$~K~km~s$^{-1}$~pc$^{2}$
from the isolated and strongly-interacting $L_{\rm{FIR}}$--$L_{\rm{CO}}$ correlations of
 \citet{solomon88}.  Therefore if we classify VII~Zw466 as a strongly-interacting
galaxy,  we appear to have observed the total $L_{\rm{CO}}$ that can be expected from the 
$L_{\rm{FIR}}$ of  VII~Zw466.  


Analogous to the $L_{\rm{FIR}}$--$L_{\rm{CO}}$ correlation, the $L_{\rm{CO}}$--$SFR$ correlation
 is often observed in samples of low-redshift star-forming galaxies \citep[e.g.\ ][]{saintonge12}.
However, the $L_{\rm{CO}}$--$SFR$ correlation is likely to be stronger since we know \
that \CO\ exists within the molecular clouds from which stars form, whereas the $L_{\rm{FIR}}$--$L_{\rm{CO}}$
 correlation is correlating the emission from the younger stellar population with the molecular gas content.  We find that the $L_{\rm{CO}}$ and $SFR$ for VII~Zw466 
is consistent with the observed $L_{\rm{CO}}$--$SFR$ correlation from \citet{saintonge12}. In 
particular, our observed $L_{\rm{CO}}$ from the LMT single pointing and the $SFR$ within the region (assuming $SFR=2.7$~\msun~year$^{-1}$) is consistent with gas depletion times between a few hundred Myr to 1~Gyr.  In this case, the inferred depletion times varies depending on the assumed  $X_{\rm{CO}}$.  However, the main result here is that the  $L_{\rm{CO}}$ and  $SFR$ of VII~Zw466 are consistent with star-forming galaxies at low redshifts \citep{saintonge12} and the general Kennicutt-Schmidt relationship.


If we assume the \Htwo\ star formation efficiency (ratio of SFR to $M_{\rm{H2}}$) to be a constant  
\citep[as is often assumed for nearby star-forming galaxies; E.g.\ ][]{leroy08,wong16},
we can estimate the \Htwo\ surface density from the star formation rate density via this expression:
\begin{equation}
\Sigma_{\rm{H2}} \approx \frac{10^{-6} \Sigma_{\rm{SFR}}}{9.51 \times 10^{-10}} \;\; {\rm{M}}_{\odot}\,{\rm{pc}}^{-2}
\end{equation}
where the star formation rate surface density ($\Sigma_{\rm{SFR}}$) is in 
units of \msun~year$^{-1}$~kpc$^{-2}$ \citep{leroy08,zheng13,wong16}.
Using the observed $SFR$ from \citet{romano08}, we find that VII~Zw466 has
an estimated $\Sigma_{\rm{SFR}}=2.8 \times 10^{-2}$~\msun~year$^{-1}$~kpc$^{-2}$ 
and $\Sigma_{\rm{H2}}=29.4$~\msun~pc$^{-2}$. Within the region observed by the 
LMT single pointing (35 percent of the total surface area estimated in Section~3.2),
 we expect $M_{\rm{H2}}=2.9 \times 10^9$~\msun, a factor of at least 1.8 greater
than that estimated from our early science LMT \CO\ observations assuming a non-Galactic
 $X_{\rm{CO}}$ conversion factor.  Such a result suggests that the star formation efficiency 
of VII~Zw466 may not be consistent with those of other nearby star-forming galaxies.



More recently, \citet{huang14}  found that the global molecular gas depletion time 
($t_{\rm{dep}} = \frac{M_{\rm{H2}}}{SFR}$) correlates strongly with
the specific star formation rate ($sSFR = \frac{SFR}{M_{\rm{stellar}}}$).
Using the total stellar mass and global star formation rate of VII~Zw466 from
Table~1, we find  log~$sSFR= -9.8$~year$^{-1}$.  Following the 
 $sSFR$--$t_{\rm{dep}}$ correlation from \citet{huang14} for a  sample
of nearby galaxies from the HERACLES survey \citep{leroy09}, we expect
log~$t_{\rm{dep}} =9.2$.  If VII~Zw466 is a star-forming galaxy that is 
comparable to the HERACLES survey of nearby star-forming galaxies, we 
would expect VII~Zw466 to have $M_{\rm{H2}}=1.2 \times 10^{10}$~\msun\ given
its observed star formation rate and stellar mass from \citet{romano08}.
In our case of VII~Zw466, the $sSFR$--$t_{\rm{dep}}$
 correlation predicts $M_{\rm{H2}}$ to be a factor of 3 to 6 greater than
the total $M_{\rm{H2}}$ estimated from our \CO\ observations.  In order to derive such a high
$M_{\rm{H2}}$ from our \CO\ observations, we would require $X_{\rm{CO}} = 9.5 \times 10^{20}$
as well as  $\Sigma_{\rm{GMC}}=17.6$~\msun\ pc$^{-2}$ using the prescription
from \citet{bolatto13}.  Such an extreme  $X_{\rm{CO}}$ suggest that 
it is unlikely that any apparent \Htwo\ deficiency of VII~Zw466 is due simply to
the assumed $X_{\rm{CO}}$.  Therefore, any apparent  \Htwo\ deficiency may either be real
or that VII~Zw466 is dominated by CO-dark gas and as such, is only \CO-deficient.
  See Section~4.3 for further discussions on CO-dark gas.

We hypothesise that a CO-deficient result is a more extreme form of  the 
``[CII]--FIR deficit'' found in ultraluminous infrared galaxies (ULIRGs) at $z=0$ where 
only approximately 10~percent of the expected [CII] luminosity is observed 
\citep[e.g.\ ][]{malhotra97,malhotra01,gracia11,narayanan16}.  The [CII] emission
originates from the outer layers of molecular clouds which are destroyed by
the strong starbursts in ULIRGs. While CO resides in more shielded regions of 
molecular clouds than [CII] \citep{narayanan16}, we posit that there is 
insufficient shielding of the CO molecules from destruction by 
in-situ star formation feedback due to the strong confinement of the galaxy's ISM in the ring.

\subsection{Atomic Hydrogen and the molecular-to-atomic Hydrogen ratio}
Atomic Hydrogen (\HI) can provide the initial gas reservoir from which molecular clouds form, 
as well as the by-product of \Htwo\ photodissociation \citep[e.g.\ ][]{allen04,higdon11}.
The \HI-normalised star formation efficiency ($SFE_{\rm{HI}} = SFR/M_{\rm{HI}}$) is surprisingly
uniform for nearby star-forming galaxies spanning across five orders of magnitude in stellar masses
 \citep[e.g.\ ][]{wong16}.  However, we find that VII~Zw466 has an 
$SFE_{\rm{HI}} = 10^{-8.72}$~year$^{-1}$.  This $SFE_{\rm{HI}}$ is greater by factors of 6 to 17 
relative to both the average $SFE_{\rm{HI}}$ found in stellar-mass selected and \HI-mass selected 
samples of star forming galaxies \citep{schiminovich10,huang12,wong16}.   Hence, the 
enhanced star formation rate  and the possible \HI\ deficiency of this
 collisional ring galaxy makes it an extreme outlier in terms of ISM and star formation
properties relative to  other non-interacting star-forming galaxies.

Even though some studies find that stars do not form solely from \Htwo\ but from cool 
gas ($T<100$~K) that are able to shield itself from the interstellar radiation field 
\citep[e.g.\ ][]{glover12,hu16}, it is commonly-thought that stars form from \Htwo\ 
\citep[e.g.\ ][]{leroy08}. Assuming that the majority of stars form from \Htwo, 
the ratio of molecular-to-atomic  Hydrogen  ($R_{\rm{mol}}$)
 reflects the efficiency with which molecular clouds form.  The $R_{\rm{mol}}$  of 
nearby galaxies have been found to  strongly depend on local conditions since the 
\HI\ is more susceptible than the \Htwo\ to tidal and ram pressure 
stripping \citep[e.g.\ ][]{chung07,vollmer12}. Processes such as tidal and ram pressure
stripping are able to remove a significant amount of \HI\ from galaxies. Indeed, 
previous \HI\ observations of VII~Zw466 by \citet{appleton96} reveal a disturbed \HI\
 morphology that bridges both VII~Zw466 and its neighbour, G2, as a result of past
tidal interactions between these two galaxies.  Such displacement of the \HI\ reservoir
 is likely to contribute to the  \HI\ deficiency
of VII~Zw466.  On the other hand, recent studies  have found that the $R_{\rm{mol}}$ in disk galaxies are
 generally governed by the hydrostatic disk pressure \citep{elmegreen89,wong16},
and easily traced by the stellar surface density \citep[$\Sigma_*$; ][]{leroy08,wong13}.  
In addition to the disk pressure and the stellar potential, stellar feedback 
  may also play a critical role in gas collapse and  molecular cloud formation 
\citep{elmegreen93,elmegreen94,hunter98,wong02,blitz04,blitz06}.

The transition between an \HI-dominated ISM and an \Htwo-dominated ISM 
occurs when $\Sigma_* =  81$~\msun~pc$^-2$, $R_{\rm{mol}} = \Sigma_* /81$~\msun~pc$^-2$ 
\citep{leroy08}. We estimate $\Sigma_* $ to be 173~\msun~pc$^{-2}$, placing 
VII~Zw466 in the \Htwo-dominated ISM regime.
Similar to other \HI-rich collisional ring galaxies 
\citep[e.g.\ ][]{parker15,higdon15,higdon11}, VII~Zw466 resides in a significant 
reservoir of \HI\ \citep{appleton96}.
In fact, the expected  $R_{\rm{mol}}$ from the estimated $\Sigma_*$ is 1.4. This 
translates to an expected $M_{\rm{H2}}=5.6 \times 10^9$~\msun,  assuming
the observed \HI\ mass of  $4.1 \times 10^9$~\msun\ \citep[see Table 1; ][]{appleton96}.

Assuming a uniform distribution of \Htwo\ across the entire galaxy, we estimate the total 
$M_{\rm{H2}}$  from our \CO\ observations to range between $1.9 \times 10^9$~\msun\ and
 $4.7 \times 10^9$~\msun\ (see Section~3.2). However, if this assumption is incorrect 
(as discussed in Section~4.1) and that our LMT observations have detected the majority
if not all the available \CO\ in VII~Zw466, then the expected $M_{\rm{H2}}$ from the 
estimated $\Sigma_*$ is likely to be a factor of 5 greater than both our single pointing
observations and those of \citet{appleton99}.

Unless a significant amount of CO in VII~Zw466 is in the form of CO-dark 
gas,  the estimated $R_{\rm{mol}}$ from our LMT observations in combination with those
of \citet{appleton99} range from 0.18 to 0.29
 depending on the assumed CO-to-\Htwo\ conversion factor.  Therefore it appears 
that the molecular gas content of collisional ring galaxies such as VII~Zw466
 are possibly deficient relative to its atomic gas content, which may also
 be deficient.  This result
is consistent with the idea of enhanced destruction of molecular 
clouds through star formation feedback within the confined ring ISM.

\subsection{CO-dark molecular gas}
So far, we have not included the contribution of CO-dark molecular gas into our
 estimation of \Htwo\ in VII~Zw466. Therefore it is possible that we have underestimated
the molecular gas reservoir in VII~Zw466. To recover the expected amount of $M_{\rm{H2}}$
  from known scaling relationships, the amount of dark CO gas will have to be at least
2.6 times that of the CO-bright gas.  

Recent high-resolution hydrodynamic simulations of the ISM in a Milky-Way analogue have found 
\CO-dark gas to exist in a diffused, low surface density state that is typically 
less shielded from the interstellar radiation field \citep[ISRF; ][]{smith14,glover16}.
 A galaxy-galaxy collision is the physical mechanism by which the high-density ring 
of VII~Zw466 formed.
However, we do not have sufficient angular resolution in the gas mapping 
(\CO\ and \HI) of this system to accurately determine the gas densities.
Therefore, we are unable to rule out the possibility for CO-dark gas in collisional
ring galaxies even though the confined high surface density ISM  within the ring
is in theory not ideal for the existence of dark CO gas.  While it is currently
beyond the scope of this paper to provide further constraints on the amount of
dark CO gas, future observations of the millimetre dust continuum \citep[e.g.\ ][]{wall16},
 higher order CO transition lines, other ionised carbon lines such as [CII] 
\citep[e.g.\ ][]{herrera15, deblok16} and high density tracers such as HCN or HNC
 can be used to help constrain the gas densities and possibly the existence of
CO-dark gas in VII~Zw466.


\subsection{Implications}

Many models \citep[e.g.\ ][]{narayanan16} make the assumption that
 $R_{\rm{mol}}$ increases as surface density of molecular clouds increases.
In general, this is confirmed by observations of both interacting and 
non-interacting star-forming galaxies in the nearby Universe.
However, our \CO\ observations and those of \citet{horellou95,appleton99} find that 
the molecular gas content of collisional ring galaxies may not be
consistent with this assumption.  The main reason for this is 
because the entire star-forming ISM of the galaxy is now confined to within 
the ring that is formed by the expanding density wave after the collision.
While we expect a temporary increase in the \Htwo\ fraction shortly
after the collision (due to compression of the ISM), the amount of \Htwo\ is 
likely to decline quickly following its consumption via the strong bursts of
star formation within the ring in the first 10~Myr \citep[e.g.\ ][]{pellerin10,henderson16}.
Alternatively, the high \Htwo\ fractions of collisional ring galaxies such
as VII~Zw466 may remain hidden due to a CO deficiency. We think that a CO 
deficiency is less likely since our observations of VII~Zw466 is consistent
with the general Kennicutt-Schmidt and the $L_{\rm{CO}}-SFR$ correlations that
are typically found from star-forming galaxies.

The stellar and ISM properties of interacting galaxies are still typically 
dominated by standard ISM conditions in regions that are unaffected or shielded from
the effects of the interactions,  in addition to the interaction-affected regions.
  These interactions (be they tidal or ram pressure)
 are typically not strong enough to alter the density of both the stellar component
and the ISM of the entire galaxy.  

Therefore, collisional galaxies represent a galactic-wide accelerated mode of 
star formation evolution whereupon the first episode of interaction-triggered 
starburst immediately has a  negative impact upon the potential for subsequent 
star formation episodes.  This argument is supported by young stellar ages 
observed in the stellar population of the ring in collisional ring galaxies
 \citep{pellerin10}.  Such extreme modes of star formation histories may be more 
important at high redshifts when  collisions are likely to be more common. 
As such, the understanding of star formation under extreme conditions
 is likely to impact our theoretical understanding of gas, 
star formation and feedback throughout the cosmic history of the Universe.

\section{Summary and conclusions}

We report the discovery of \CO\ in the ring galaxy VII~Zw466 using the 
RSR instrument during the early science commissioning phase of the LMT. 
 We find the CO luminosity ($L_{\rm{CO}}$) of VII~Zw466 to be 
$1.6 \times 10^8$~K~km~s$^{-1}$~pc$^{2}$.

Recent studies have found that \CO\ observations of collisional ring galaxies
 are likely to underestimate the true amount of \Htwo\ using the standard Galactic  
$X_{\rm{CO}}$ value.  In this paper, we test two different prescriptions 
for converting between the observed \CO\ emission to the inferred  $M_{\rm{H2}}$ 
from \citet{narayanan12} and \citet{bolatto13}.  Using these two  prescriptions, 
we find that the estimated \Htwo\ mass for VII~Zw466 is a factor of 0.9 to 1.9 times greater 
than that found from the standard Solar neighbourhood $X_{\rm{CO}}$ value. The 
estimated $M_{\rm{H2}}$ for VII~Zw466 from our LMT observations ranges between 
$7.5 \times 10^8$~\msun\ and $1.6 \times 10^9$~\msun.  

If we assume a uniform \Htwo\ distribution and scale the observed 
$M_{\rm{H2}}$ to account for the entire surface area of VII~Zw466, 
our estimated total $M_{\rm{H2}}$ ranges between 
$1.9 \times 10^9$~\msun\ and $4.7 \times 10^9$~\msun. However, we think
that such an assumption may not be valid since previous observations by \citet{appleton99} 
have found an $M_{\rm{H2}}$ upper limit of $<1.2 \times 10^9$~\msun\ for this galaxy.
Therefore, we posit that our LMT \CO\ observations have detected the majority
of the \CO\ content in this galaxy and that the molecular gas is not likely
to be evenly distributed across the entire galaxy.


We conclude that collisional ring galaxies represent the 
extreme end of galaxy interactions and evolution whereby 
 the ISM and star forming properties are distinct from 
other star-forming galaxies that are isolated or experiencing 
non-collisional interactions. Our results suggest that the 
an  \Htwo\ deficiency in the star-forming environment of the ring is likely 
real unless there exists a significant amount of CO-dark gas. 
Also the apparent \Htwo\ deficiency  is unlikely to be explained by 
the variation of $X_{\rm{CO}}$ from the standard Galactic value.

\vspace{1cm}

\chapter{\bf{Acknowledgments.}}
We would like to thank the LMT observatory staff for all their help
and advice with the observations  at the LMT.  
This research has made use of the NASA/IPAC Extragalactic Database
(NED) and the NASA/IPAC Infrared Science Archive (IRSA), which is operated by the 
Jet Propulsion Laboratory, California Institute of Technology, under contract with 
the National Aeronautics and Space Administration.

\bibliographystyle{mnras} 
\bibliography{mn-jour,paperef}
\end{document}